%% file: btech_thesis_template.tex
\documentclass[letterpaper,11pt]{report}

\usepackage{fullpage}
\usepackage{verbatim}

\usepackage{setspace}
\usepackage{fancyhdr}
\usepackage{xcolor}
\usepackage{biblatex}
\usepackage[small]{caption}
\usepackage{graphics}
\usepackage{color}
\usepackage{subcaption}
\usepackage{hyperref}

\usepackage[dvips]{graphicx}

\def\title{Multiple Myeloma Cancer Cell Instance Segmentation}

\begin{document}

\include{btech_thesis_cover_dikshant}

%

\pagebreak

\begin{abstract}
Images remain the largest data source in the field of healthcare. But at the same time, they are the most difficult to analyze. More than often, these images are analyzed by human experts such as pathologists and physicians. But due to considerable variation in pathology and the potential fatigue of human experts, an automated solution is much needed. The recent advancement in Deep learning could help us achieve an efficient and economical solution for the same. In this research project, we focus on developing Deep Learning based solution for detecting Multiple Myeloma cancer cells using an Object Detection and Instance Segmentation System. We explore multiple existing solutions and architectures for the task of Object Detection and Instance Segmentation and try to leverage them and come up with a novel architecture to achieve comparable and competitive performance on the required task. To train our model to detect and segment Multiple Myeloma cancer cells, we utilise a dataset curated by us using microscopic images of cell slides provided by Dr.Ritu Gupta(Prof., Dept.  of Oncology AIIMS).

\vspace{2in}
\end{abstract}

\newpage

\section*{Acknowledgments}\label{section:acknowledgments}
\pagestyle{plain}
\pagenumbering{roman}

I thank my advisor, Dr. Anubha Gupta, for giving me an opportunity to work under her guidance. Her advice and constant support have enabled me to think critically about the domain of research and learn to come up with concise, effective solutions to the practical problem.

I thank Dr.Ritu Gupta(Prof., Dept. of Oncology AIIMS), for providing a large databank of patient cell slide images, which helped us create a benchmark dataset in the field of Multiple Myeloma Cancer Detection.

I also thank Mr.Shubham Goswami(PhD, IIIT Delhi) for his immense contribution to the project. His vast knowledge of the various methods pertaining to Bio-Medical Computer Vision and readiness to brainstorm solutions for the challenges along the way have been a significant driver to this project.

Finally, I thank Mr. Bharat Giddwani (Deep Learning Solutions Architect, NVIDIA) for providing access to DGX Workstations for us to be able to train large scale deep learning models.

\vspace{2in}
\section*{Work Distribution}

Monsoon Semester 2020 (4 credits)
\begin{enumerate}
    \item Dataset: Annotations and Auxiliary Dataset Exploration (Sept - Nov, 2020)
    \item Existing Deep Learning Architectures: Literature Survey (Nov - Dec, 2020)
    \item Proposed Architecture: Backbone Network (Dec, 2020)
\end{enumerate}

Winter Semester 2021 (6 Credits)
\begin{enumerate}
    \item Proposed Architecture: Feature Network (Jan - Feb, 2021)
    \item Proposed Architecture: Heads (Feb, 2021)
    \item Experiments and Results:  (March - April, 2021)
\end{enumerate}

\newpage

\tableofcontents



\newpage
\mbox{}


\chapter{Introduction}\label{chapter:introduction}
One of the major challenges in developing nations like India is to make medical facilities accessible to the entire population of the country. Due to a lack of medical infrastructure throughout the country, A significant percentage of the country’s population faces financial and physical outreach to specialized hospitals and treatments. Cancer detection is one of the major life-threatening health-related hazards and even a small percentage of the population getting diagnosed with cancer can result in large numbers given India’s population. There are specialized hospitals being developed to address this need but in the present situation much of the population cannot afford the expensive diagnostics like flow cytometry test, Also much of these specialized hospitals are present or being laid out in urban areas due to which people available in remote areas are not able to avail these facilities.  \\

Artificial intelligence(AI) based solutions can prove themselves as no less than a miracle in such scenarios. The advances in Deep Learning(DL a way to achieve AI) has increased the possibilities to use DL based solutions in the Medical domain and for diagnostics. DL-based solutions show more generalisability on a prospective subject’s data by learning the distinguishable characteristics from the data itself rather than relying on handcrafted features that are used by traditional statistics. Even though DL based solutions have shown their potential in many medical applications like diagnostics and are becoming the state of the art solutions. There are many fundamental challenges related to DL based solutions like explainability, capturing heterogeneity at various levels(subject, sensor, site), label noise, handling uncertainty, etc which need to be addressed for their usage as an alternative to existing diagnostics. \\

This BTP address some of the above described fundamental challenges in DL based solutions in the application of white blood cancer diagnostics, Especially Multiple Myeloma(MM) cancer diagnostics. For this, We’ll be exploring and developing DL based solutions for the detection and segmentation of MM cancer cells. This research work is being carried out in collaboration with the Dept. Of Oncology, AIIMS under the supervision and expert guidance of Dr. Anubha Gupta(Prof., Dept. of ECE, IIIT-Delhi) and Dr.Ritu Gupta(Prof., Dept. of Oncology AIIMS). The outcomes of this research include aiming for publications for validation of our research followed by deployment of the proposed DL based solution in AIIMS(Delhi).

\pagenumbering{arabic}
\setcounter{page}{1}
\onehalfspacing

\chapter{Dataset} \label{chapter:Dataset}
\section{Primary Dataset}
\subsection{Overview}
\begin{figure}[!h]
\centering
     \begin{subfigure}[b]{0.3\textwidth}
         \centering
         \includegraphics[width=\textwidth]{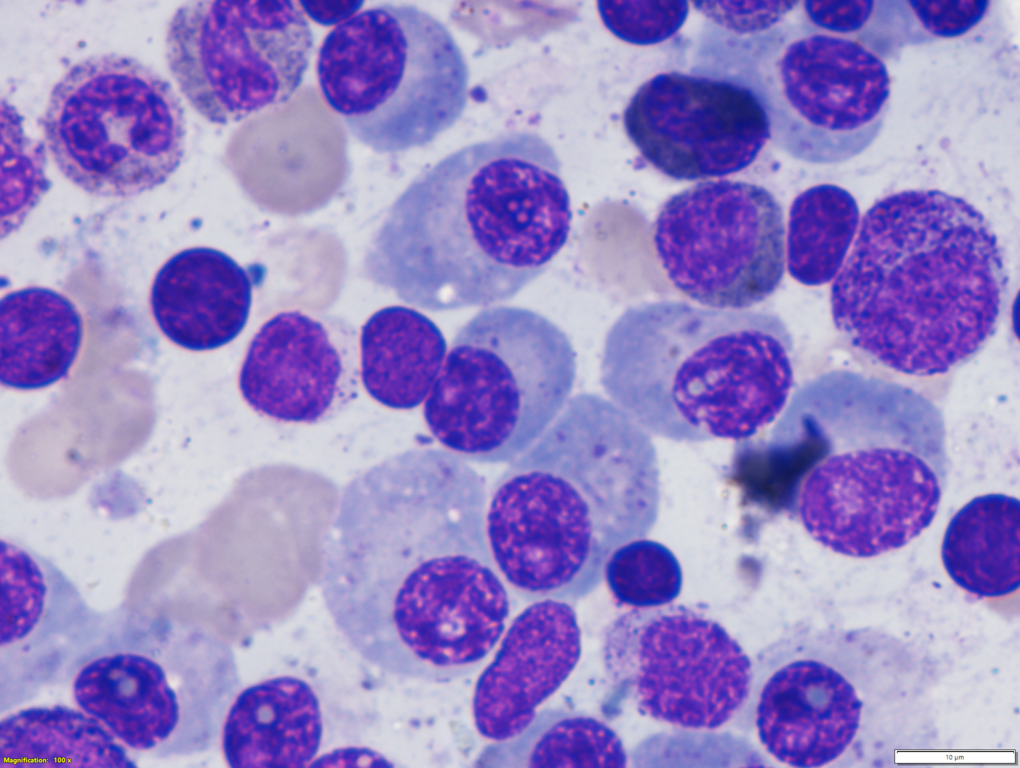}
         \caption{Stained Cell Slide Image}
         \label{fig:y equals x}
     \end{subfigure}
     \hfill
     \begin{subfigure}[b]{0.3\textwidth}
         \centering
         \includegraphics[width=\textwidth]{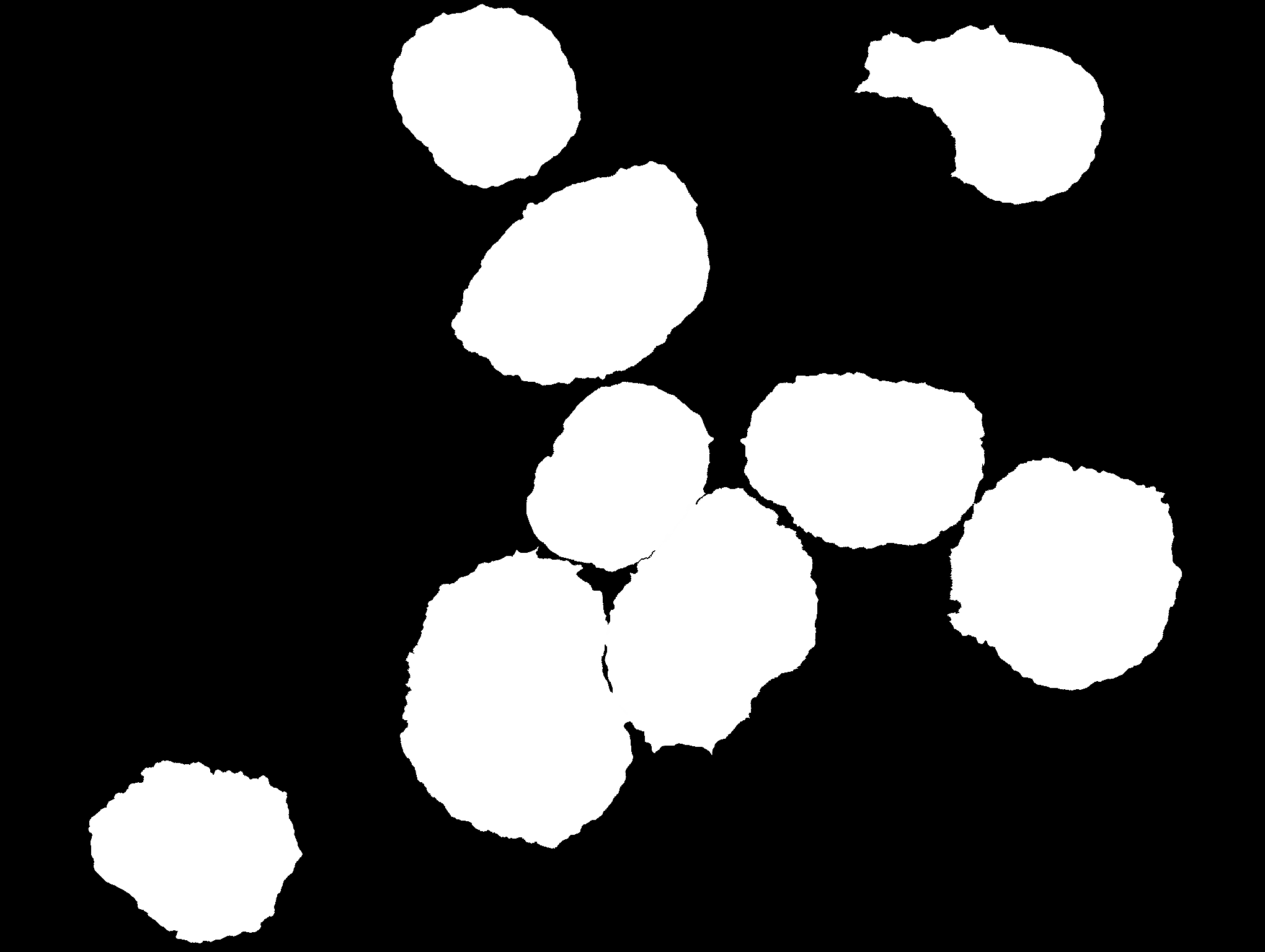}
         \caption{Annotated Masks}
         \label{fig:three sin x}
     \end{subfigure}
     \hfill
     \begin{subfigure}[b]{0.3\textwidth}
         \centering
         \includegraphics[width=\textwidth]{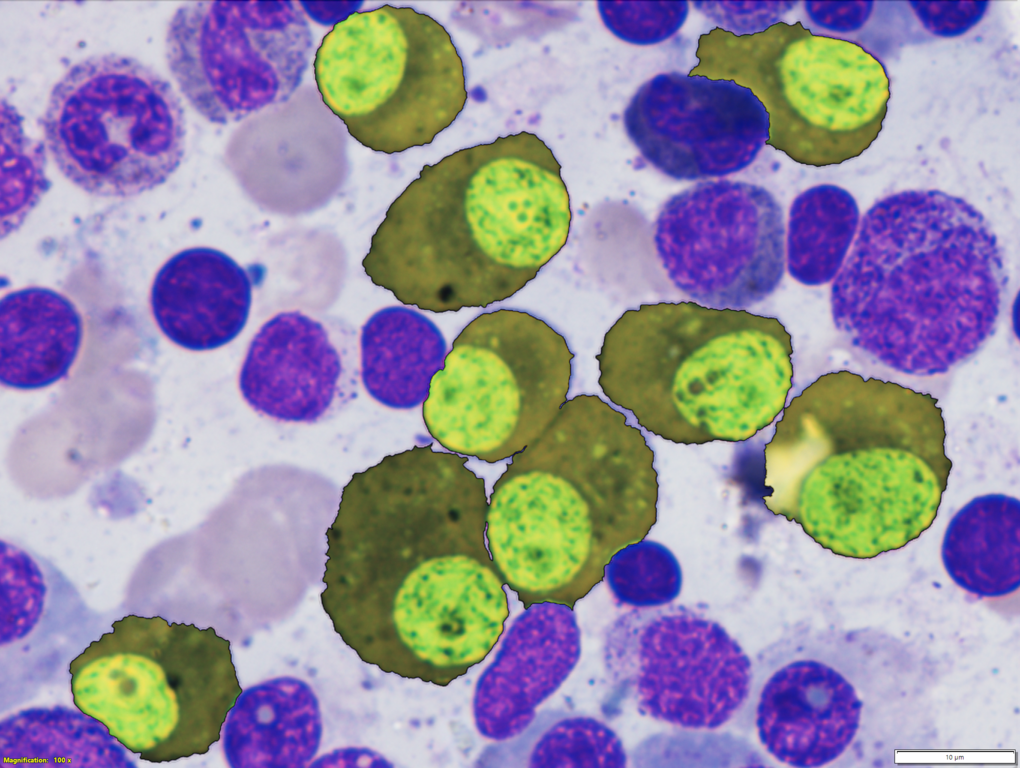}
         \caption{Masks Overlayed}
         \label{fig:five over x}
     \end{subfigure}
\caption{Our Dataset}
\label{fig:data fig}
\end{figure}
Our primary dataset is a set of microscopic images that were captured from the bone marrow aspirate slides of multiple subjects, as per standard guidelines\cite{palumbo2014international}.
The data is compiled from the bone marrow aspirates slides of n subjects and consists of 35,500 images with over 1,50,000 cell markers.
The slides were stained using the Jenner-Giemsa stain. The images were captured using a Nikon Eclipse-200 microscope equipped with a digital camera at a magnification of 1000x. These images were captured in a raw BMP format with a size 2040 × 1536 pixels. 

\begin{figure}[!h]
\centering
         \includegraphics[width=\textwidth]{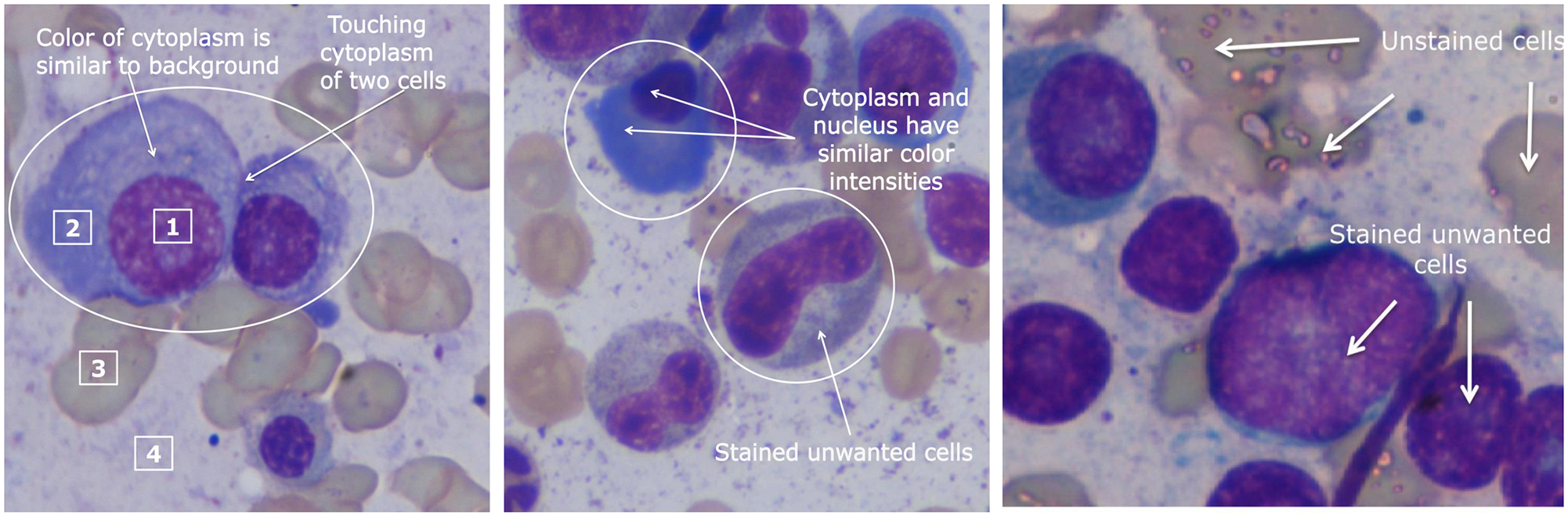}
\caption{Challenges Posed \cite{gupta2018pcseg}}
\label{fig:probelems fig}
\end{figure}

The given dataset poses the following challenges:
\begin{enumerate}
  \item The task of segmentation required us to segment both the nucleus and the cytoplasm. For some cells, the cytoplasm may be in less contrast to the nucleus or at times, even the background. This causes difficulty to segment out the cytoplasm.
  \item Presence of more than one type of stained and unstained cells poses another challenge in extracting plasma cells of interest
  \item Plasma cells may be clustered together and hence, segmentation of the overlapping/touching cells is required. Generally, this is difficult because of different configurations:
  \begin{itemize}
     \item nuclei of different cells are touching
     \item nuclei of one and cytoplasm of another cell are touching
     \item the cytoplasm of different cells are touching
   \end{itemize}
\end{enumerate}

A subset of the dataset previously released is available at the public repository\cite{gupta2018pcseg}. 
\begin{figure}[!h]
\centering
         \includegraphics[width=0.7\textwidth]{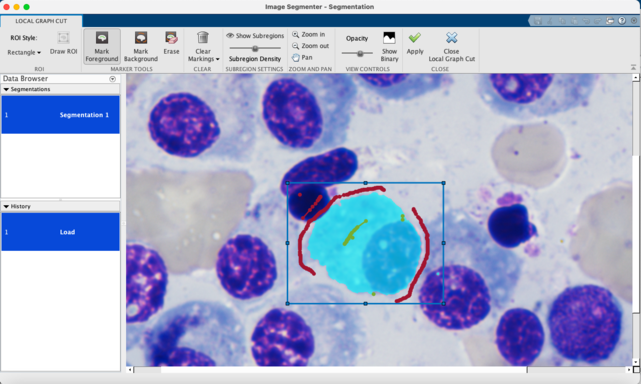}
         
\caption{MATLAB Image Segmentor on Multiple Myeloma Images}
\label{fig:segmenter_matlab fig}
\end{figure}
\subsection{Annotation}
The Obtained data consisted of around 35,500 images of stained Multiple Myeloma plasma cells, with each image having multiple markers to be annotated ranging from 0-20. The data was segmented into batches of 100 images each giving around 350 batches in total to be annotated. Students doing projects under Dr. Anubha Gupta on Multiple Myeloma were distributed batches and were tasked to annotate masks(See Fig. \ref{fig:data fig}) for the markers on these images. A proper NDA was signed to maintain data and patient confidentiality.

We used the Image Segmenter Tool from MATLAB\cite{matlab-simulink} to achieve the annotated masks.(See Fig.\ref{fig:segmenter_matlab fig})

We use a subset of this dataset along with previous similar dataset from SBIlab at IIIT Delhi to create a dataset, which was also made a public challenge in SegPC-2021. This subset contains 775 images. These are divided into a training set of 298 images, validation set of 200 images, and a test set of 277 images. For this BTP's purpose we use the train set for training and validation set for performance metric purposes. 

\section{Auxiliary Dataset}

\subsection{MS COCO}
\begin{figure}[!h]
\centering
         \includegraphics[width=\textwidth]{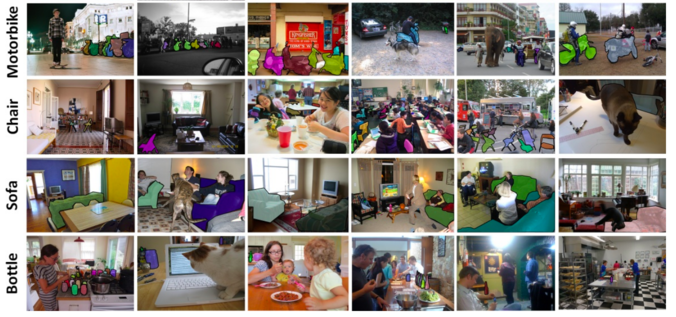}
         
\caption{Example from COCO Dataset}
\label{fig:coco_eg fig}
\end{figure}
\begin{figure}[!h]
\centering
         \includegraphics[width=\textwidth]{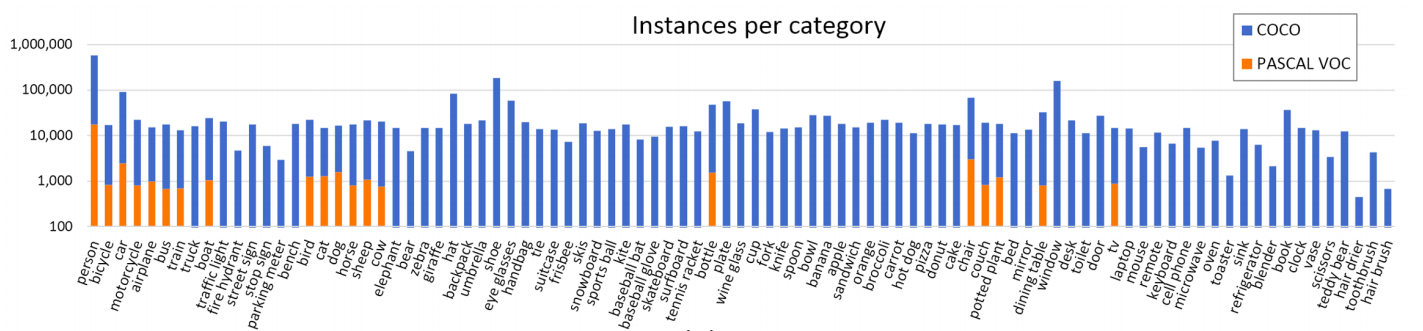}
         
\caption{Comparitive class distribution of MSCOCO w.r.t PASCAL VOC \cite{lin2014microsoft}}
\label{fig:coco_dist fig}
\end{figure}

The MS COCO dataset\cite{lin2014microsoft} is a large-scale dataset of 328k images consisting of over 2.5 million objects. The dataset was curated by gathering images of complex everyday scenes containing common objects in their natural context. Objects were labeled using per-instance segmentations to aid in
precise object localization. The comparitive class distribution of the COCO dataset with respect to the PASCAL VOC dataset is shown in Fig \ref{fig:coco_dist fig}

\chapter{Existing Deep Learning Architectures}\label{chapter:Existing Deep Learning Architectures}
We explored State of the art architectures on some of the benchmark datasets for Object Detection and Instance Segmentation namely MS COCO and  PASCAL VOC 2007 in the non-medical domain. 

Following are some of the recent advancements in Object Detection and Instance Segmentation Architectures :

\section{EfficientDet: Scalable and Efficient Object Detection}
\begin{figure}[!h]
    \centering
    \includegraphics[width=1.0\linewidth]{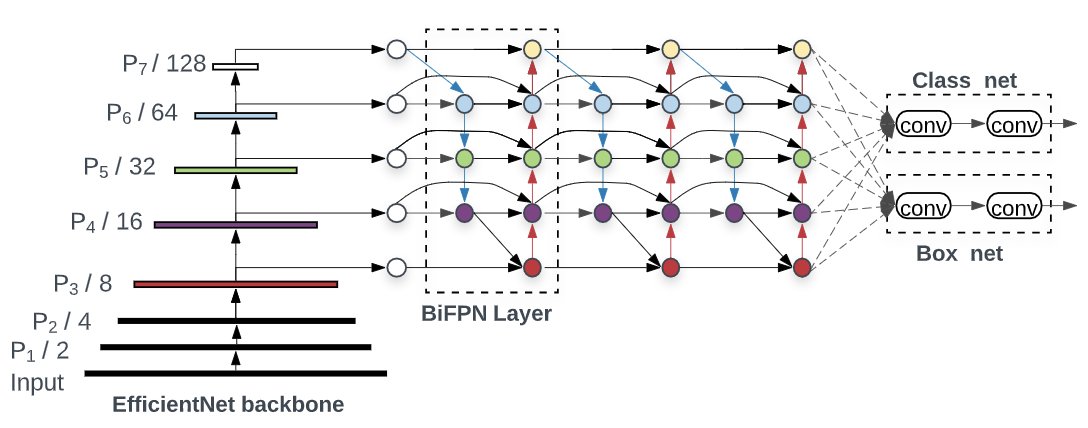}
    \caption{EfficientDet Model Architecture \cite{tan2020efficientdet}}
    \label{fig:effdet model}
\end{figure}

Existing object detectors are mostly categorized by whether they have a region-of interest proposal step (two-stage) or not (one stage). While two-stage detectors tend to be more flexible and more accurate, one-stage detectors are often considered to be simpler and more efficient by leveraging predefined anchors. EfficientDet \cite{tan2020efficientdet} largely follows the one-stage detectors paradigm.It employs ImageNet-pretrained EfficientNets \cite{tan2020efficientnet} as the backbone network. Their proposed BiFPN\cite{tan2020efficientdet} serves as the feature network, which takes level 3-7 features \{P3, P4, P5, P6, P7\} from the backbone network and repeatedly applies top-down and bottom-up bidirectional feature fusion. These fused features are fed to a class and box network to produce object class and bounding box predictions respectively (See Fig. \ref{fig:effdet model}). Similar to RetinaNet \cite{lin2018focal}, the class and box network weights are shared across all levels of features.

\section{CSPNet: A New Backbone that can Enhance Learning Capability of CNN}
\begin{figure}[!h]
    \centering
    \includegraphics[width=1.0\linewidth]{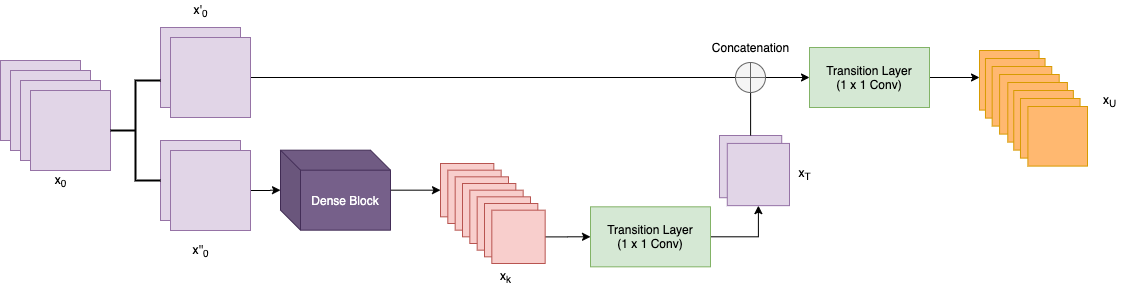}
    \caption{CSP-DenseNet Block Architecture}
    \label{fig:csp model}
\end{figure}

This Paper introduces a concept of partial skip connections in repetitive blocks of a backbone network. They suggest that rather than passing the whole feature map through a block and also forward on a skip connection, we pass half the channels of the feature map through the block and half through the skip-connection (See Fig.\ref{fig:csp model}).This partitioning of the feature map helps achieve a richer gradient combination due to gradient flow propagating through different network paths while reducing the amount of computation. The proposed CSPNet\cite{wang2019cspnet} can be easily applied to ResNet\cite{he2015deep}, ResNeXt\cite{xie2017aggregated}, and DenseNet\cite{huang2018densely}. After applying CSPNet on the above mentioned networks, the computation effort can be reduced from 10\% to 20\%, but it outperforms in terms of accuracy, in conducting image classification task on ImageNet\cite{russakovsky2015imagenet}.

\begin{figure}[!h]
    \centering
    \includegraphics[width=0.5\linewidth]{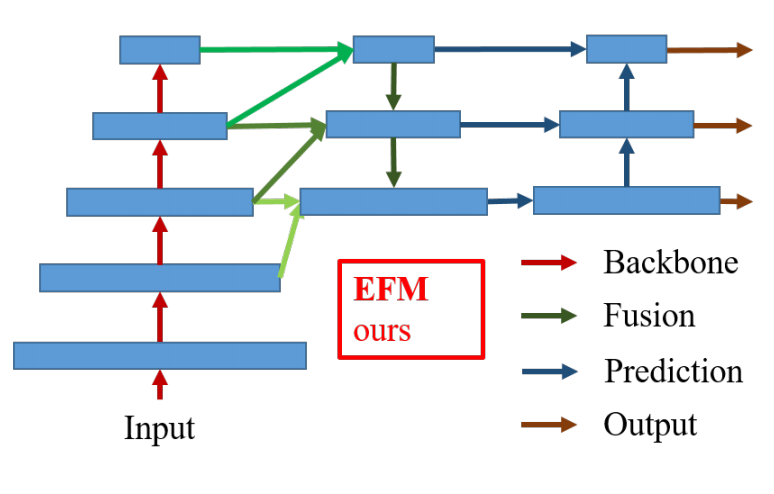}
    \caption{Exact Fusion Model \cite{wang2019cspnet}}
    \label{fig:efm model}
\end{figure}

They also propose a feature aggregation network called the Exact Fusion Model\cite{wang2019cspnet} inspired from YOLOv3\cite{redmon2018yolov3} which assigns exactly one bounding-box prior to each ground truth object.Each ground truth bounding box corresponds to one anchor box that surpasses the threshold IoU. If the size of an anchor box is equivalent to the field-of-view of the grid cell, then for the grid cells of the s\textsuperscript{th} scale, the corresponding bounding box will be lower bounded by the (s - 1)\textsuperscript{th} scale and upper bounded by the (s + 1)\textsuperscript{th} scale. Therefore, the EFM assembles features from the three scales. See Fig.\ref{fig:efm model}

\section{SpineNet: Learning Scale-Permuted Backbone for Recognition and Localization}
\begin{figure}[!h]
\centering
     \begin{subfigure}[b]{0.5\textwidth}
         \centering
         \includegraphics[width=\textwidth]{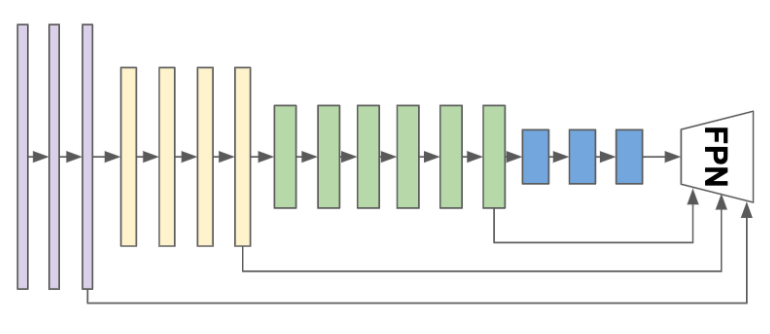}
         \caption{ResNet-50-FPN \cite{he2015deep}}
         \label{fig:resnet-50}
     \end{subfigure}
     \hfill
     \begin{subfigure}[b]{0.5\textwidth}
         \centering
         \includegraphics[width=\textwidth]{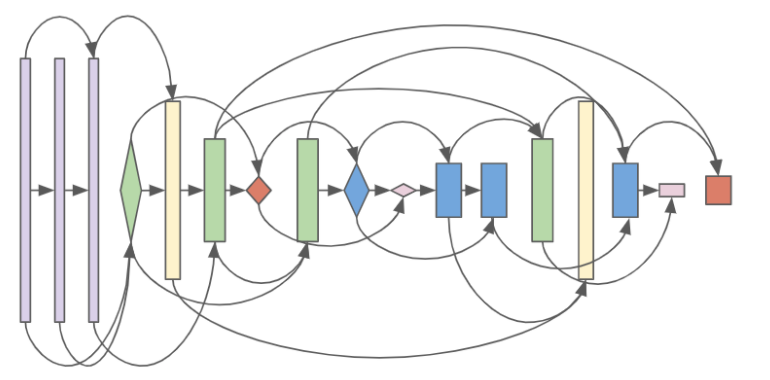}
         \caption{SpineNet-49 \cite{du2020spinenet}}
         \label{fig:spinenet-49}
     \end{subfigure}
\caption{The architecture comparison between a ResNet backbone (\ref{fig:resnet-50}) and the SpineNet backbone (\ref{fig:spinenet-49}) derived from it using NAS. }
\label{fig:spinenet fig}
\end{figure}

CNNs created for image tasks typically encode an input image into a sequence of intermediate features that capture the semantics of an image (from local to global), where each subsequent layer has a lower spatial dimension. However, this scale-decreased model may not be able to deliver strong features for multi-scale visual recognition tasks where recognition and localization are both important (e.g., object detection and segmentation).
Several works including FPN\cite{lin2017feature} and DeepLabv3+\cite{chen2018encoderdecoder} propose multi-scale encoder-decoder architectures to address this issue, where a scale-decreased network (e.g., a ResNet) is taken as the encoder (commonly referred to as a backbone model). A decoder network is then applied to the backbone to recover the spatial information.

While this architecture has yielded improved success for image recognition and localization tasks, it still relies on a scale-decreased backbone that throws away spatial information by down-sampling, which the decoder then must attempt to recover.

They propose a meta architecture called a scale-permuted model that enables two major improvements on backbone architecture design. First, the spatial resolution of intermediate feature maps should be able to increase or decrease anytime so that the model can retain spatial information as it grows deeper. Second, the connections between feature maps should be able to go across feature scales to facilitate multi-scale feature fusion. They then use neural architecture search (NAS)\cite{zoph2017neural} with a novel search space design that includes these features to discover an effective scale-permuted model.

They named the learned 49-layer scale-permuted backbone architecture SpineNet-49\cite{du2020spinenet} from the ResNet-50\cite{he2015deep} model. SpineNet-49 can be further scaled up to SpineNet-96/143/190 by repeating blocks two, three, or four times and increasing the feature dimension. An architecture comparison between ResNet-50-FPN and the final SpineNet-49 is shown.(See Fig. \ref{fig:spinenet fig})

\chapter{Proposed Architecture}\label{chapter:Proposed Architecture}
\begin{figure}[!h]
    \centering
    \includegraphics[width=1.0\linewidth]{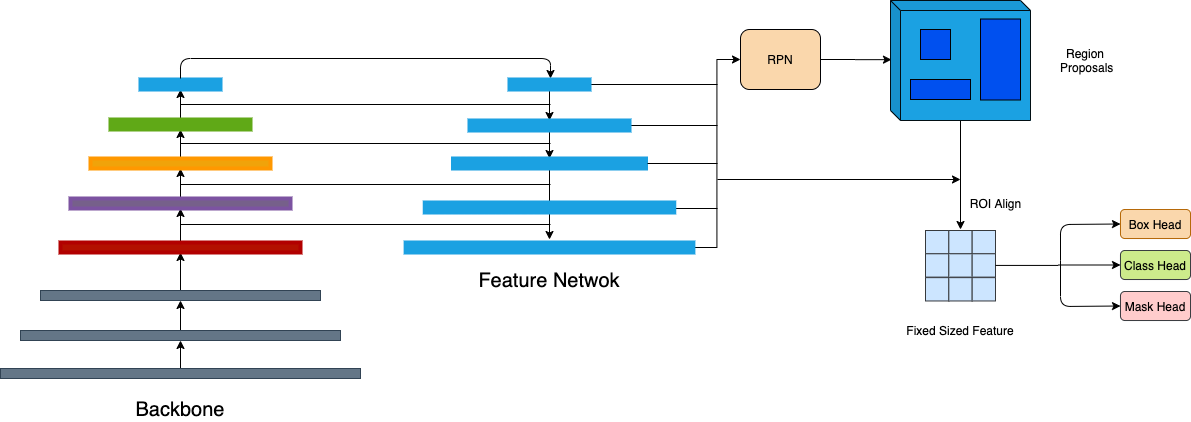}
    \caption{Two Step Object Detection - Segmentation Model}
    \label{fig:two-step model}
\end{figure}
We model our proposed novel architecture on a two-step Object Detection - Segmentation Architecture(See Figure \ref{fig:two-step model}) based on the Mask-RCNN\cite{he2018mask}. The main reason for this choice is that though one-step models are known to be faster in inference time, they lack at the accuracy metric when compared to two-step models\cite{Jiao_2019}. And our specific problem of detecting cancer cells require accuracy to be the top most-priority and not the inference time.

\section{Backbone}
\subsection{CSP + EfficientNet + SAM }
\begin{figure}[!h]
    \centering
    \includegraphics[width=0.7\linewidth]{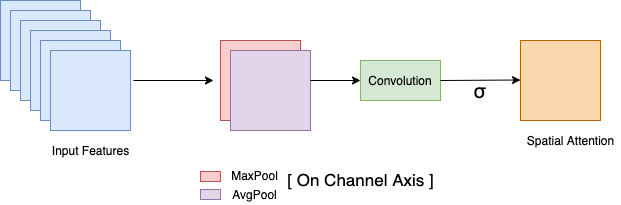}
    \caption{Spatial Attention Module}
    \label{fig:SAM module}
\end{figure}

One current candidate for the backbone that we have devised is a modified EfficientNet\cite{tan2020efficientnet} with CSP Connections\cite{wang2019cspnet}(See Section 3.2) and a Spatial Attention Module(SAM)\cite{woo2018cbam} (See Fig.\ref{fig:SAM module}) in the repetitive MBConv Block(Inverted Residual Block)\cite{sandler2019mobilenetv2} of the original efficientnet. We found this configuration and architecture to give much better performance on CIFAR10\cite{Krizhevsky2009LearningML} Classification Task and due to the CSP Connections the FLOPs are reduced and the architecture remains both scalable and efficient. We call our backbone architectures a CSP-EfficientNet Family of models with variants same as EfficientNets\cite{tan2020efficientnet} ranging from B0 to B7.

\section{Feature Network}
\subsection{NAS-FPN}
\begin{figure}[!h]
    \centering
    \includegraphics[width=0.7\linewidth]{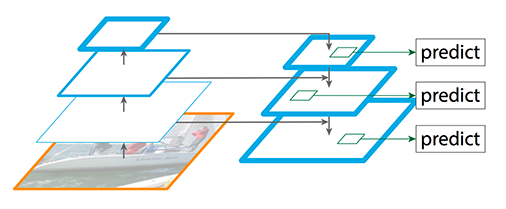}
    \caption{FPN\cite{lin2017feature}}
    \label{fig:fpn}
\end{figure}

 Feature networks are feature extractors and recent state of the art fetaure networks include FPN(Feature Pyramid Network)\cite{lin2017feature}.FPN composes of a bottom-up and a top-down pathway. The bottom-up pathway is the usual convolutional network for feature extraction. As we go up, the spatial resolution decreases. With more high-level structures detected, the semantic value for each layer increases. FPN also provides top-down pathway to construct higher resolution layers from the semantic rich layers with lateral connections between reconstructed layers and the corresponding feature maps to help the detector to predict the location betters (See Fig. \ref{fig:fpn}).
 
\begin{figure}[!h]
    \centering
    \includegraphics[width=1.0\linewidth]{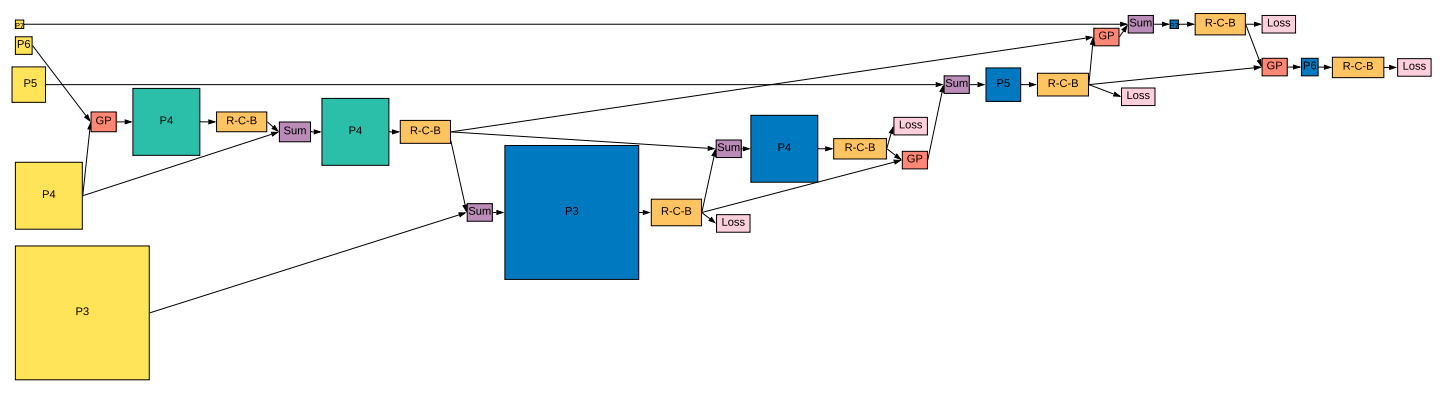}
    \caption{Architecture of the discovered 7-merging-cell pyramid network in NAS-FPN with 5 input layers (yellow) and 5
output feature layers (blue). GP and R-C-B are stands for Global Pooling and ReLU-Conv-BatchNorm, respectively.\cite{ghiasi2019nasfpn}}
    \label{fig:nasfpn}
\end{figure}

 For the feature network in our model we incorporate a novel architecture known as NAS-FPN\cite{ghiasi2019nasfpn} that is discovered using neural architecture search. The search is a novel scalable search space covering all cross-scale connections possible in a vanilla FPN\cite{lin2017feature}. The discovered architecture consists of a combination of top-down and bottom-up connections to fuse features across scales. The optimal intra-scale connections found through the Neural architecture search are visualised in Fig.\ref{fig:nasfpn}
 
\section{Heads}
\subsection{Mask Head : Dice Loss Objective}

Instead of the normal Sigmoid Crossentropy loss on the mask head of the MaskRCNN type model, we incorporate a new loss using the Dice Coefficient\cite{Dice_1945} which in theory will tend to maximise the mean IOU of the predicted masks and the ground truth masks of the samples by minimising the negative of the Dice Coefficient. The following is the expression for the dice coefficient :

\begin{equation}
    Dice = \frac{2|X \cap Y|}{|X| + |Y|} 
\end{equation}
inspired from \cite{Dice_1945},
where X is the predicted mask and Y is the ground truth mask. Then our dice loss is given by: 
\begin{equation}
    Dice Loss = -1 * Dice
\end{equation}

\chapter{Experimental Results}\label{chapter:Experimental Results}
\section{Backbone}
\begin{figure}[!h]
\centering
     \begin{subfigure}[b]{0.3\textwidth}
         \centering
         \includegraphics[width=\textwidth]{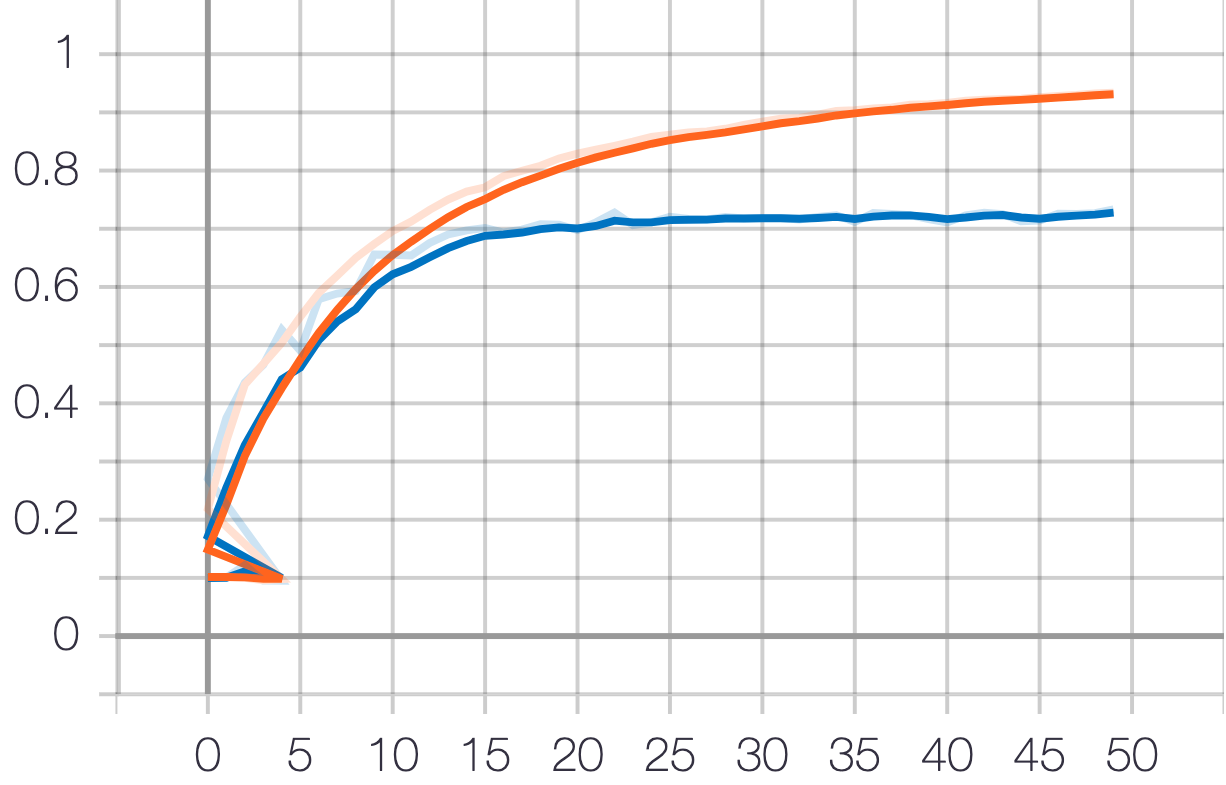}
         \caption{EfficientNetB0}
         \label{fig:eff}
     \end{subfigure}
     \hfill
     \begin{subfigure}[b]{0.3\textwidth}
         \centering
         \includegraphics[width=\textwidth]{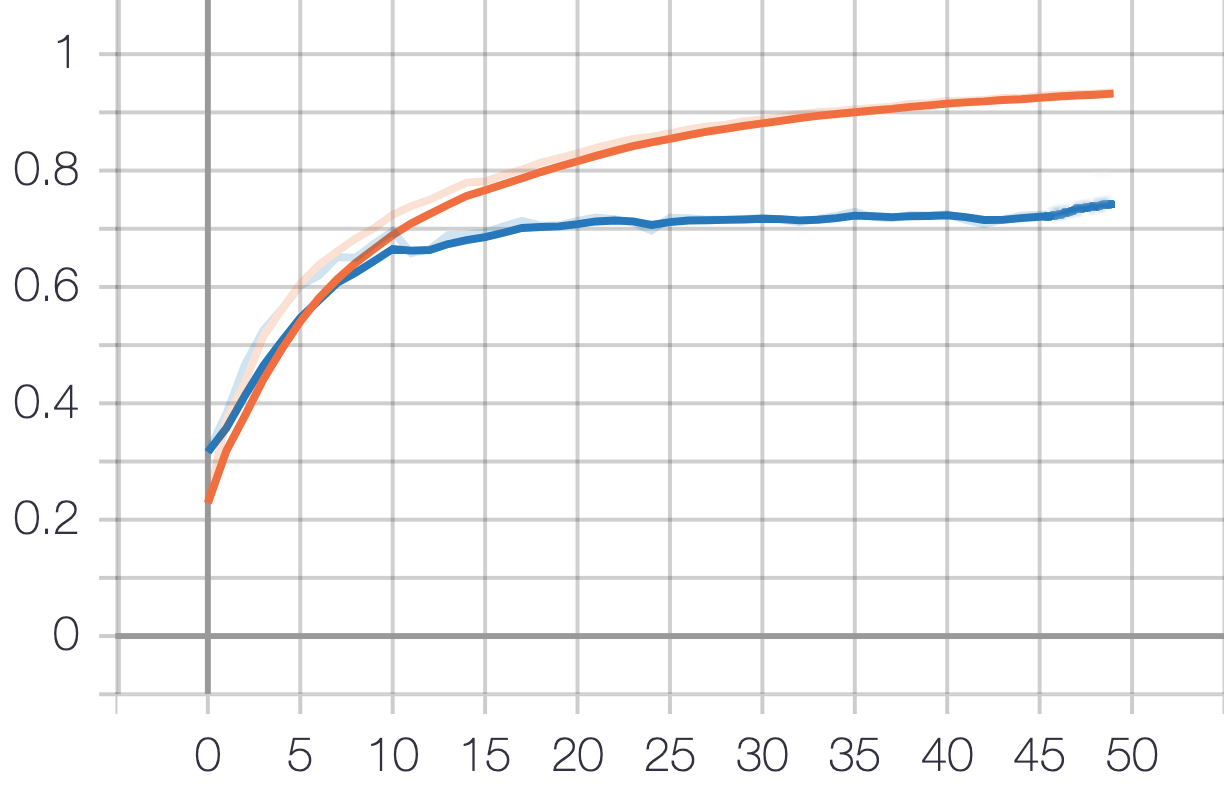}
         \caption{EfficientNetB0 + SAM}
         \label{fig:effsam}
     \end{subfigure}
     \hfill
     \begin{subfigure}[b]{0.3\textwidth}
         \centering
         \includegraphics[width=\textwidth]{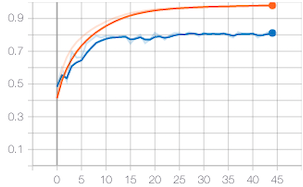}
         \caption{EfficientNetB0 + SAM + CSP}
         \label{fig:effsamcsp}
     \end{subfigure}
\caption{Classification accuracy plots for experimented architectures.}
\label{fig:plot fig}
\end{figure}

For Backbone experimentation purposes we took EfficientNetB0\cite{tan2020efficientnet} as the base architecture and CIFAR10\cite{Krizhevsky2009LearningML} as the metric dataset to compare performances which is in accordance to the norm in computer vision classification task before training on a larger dataset like Imagenet\cite{russakovsky2015imagenet}. We fixed training hyperparamters to be same for every architecture and epochs for the architectures to 50, as we observed complex architectures like these converged relatively faster on a small dataset like CIFAR10\cite{Krizhevsky2009LearningML} which can be seen in Fig.\ref{fig:plot fig} on \colorbox{orange}{Train} and \colorbox{blue}{Validation} data. The accuracy results can be seen in Table \ref{table:1}

\begin{table}[!h]
\centering
\caption{Val. accuracy of experimented backbone architectures after 50 epochs on CIFAR10\cite{Krizhevsky2009LearningML}  classification task.}
\label{table:1}
\begin{tabular}{ |c|c| } 
 \hline
 \textbf{Architecture} & \textbf{Val. Accuracy}  \\ 
 \hline
 EfficientNetB0 & 0.71  \\ 
 \hline
 EfficientNetB0 + Spatial Attention & 0.75  \\ 
 \hline
 EfficientNetB0 + Spatial Attention + CSP & \textbf{0.82}  \\ 
 \hline
    
\end{tabular}
\end{table}

\section{Feature Network}

To understand the contribution of the NAS-FPN\cite{ghiasi2019nasfpn} feature network, we perform ablation study on the performance of the model on the COCO dataset\cite{lin2014microsoft}.

\begin{table}[!h]
\centering
\caption{Performace of our Model with NAS-FPN\cite{ghiasi2019nasfpn} and other feature networks on COCO minival\cite{lin2014microsoft}.}
\label{table:2}
\begin{tabular}{ |c|c|c| } 
 \hline
 \textbf{Architecture} & \textbf{BoxAP} & \textbf{MaskAP}  \\ 
 \hline
 CSP-EfficientnetB0 + C4 & 40.3 & 35.3 \\ 
 \hline
 CSP-EfficientnetB0 + FPN & 42.3  & 37.2 \\ 
 \hline
 CSP-EfficientnetB0 + NAS-FPN & \textbf{ 43.8} & \textbf{38.3} \\ 
 \hline
    
\end{tabular}
\end{table}

\section{Heads}
For our model we train our mask head with a loss function based on IoU score called the dice loss , unlike the conventional Mask-RCNN's Mask head which trains on just sigmoid binary crossentropy of the predicted binary masks from the GT masks. So we perform ablation study to understand the contribution of the modified loss objective of the mask head to the performance of the whole model.

\begin{table}[!h]
\centering
\caption{MaskAP of our model(CSP-EfficientNetB0 + NAS-FPN) with and without the modified loss objective on the COCO minival\cite{lin2014microsoft}.}
\label{table:3}
\begin{tabular}{ |c|c|} 
 \hline
 \textbf{Mask Head Loss} & \textbf{MaskAP}  \\ 
 \hline
 Sigmoid Binary Crossentropy Loss & 38.3  \\ 
 \hline
 Dice Loss & \textbf{39.1}  \\ 
 \hline

\end{tabular}
\end{table}

\section{Our Model}

In this section we show the performance of our model in comparison with other existing comparable architectures on the same task.

\begin{table}[!h]
\centering
\caption{Performance of our model in comparison with comparable existing MaskRCNN\cite{he2018mask} model architectures on the COCO minival\cite{lin2014microsoft}.}
\label{table:4}
\begin{tabular}{ |c|c|c|c| } 
 \hline
 \textbf{Work} & \textbf{Model} & \textbf{BoxAP} & \textbf{MaskAP}  \\
 \hline
        & ResNet50-FPN\cite{wu2019detectron2} & 38.6 & 35.2  \\ 
 Prior  & SpineNet-49 & 41.2 & 36.3  \\ 
        & EfficientNetB0-FPN & 41.4 & 36.7  \\ 
 
\hline
 \textbf{Ours} & \textbf{CSP-EfficientNetB0 + NAS-FPN + Dice Loss }& \textbf{43.8} & \textbf{39.1}  \\ 
 \hline

\end{tabular}
\end{table}

\begin{figure}[!h]
    \centering
    \includegraphics[width=1.0\linewidth]{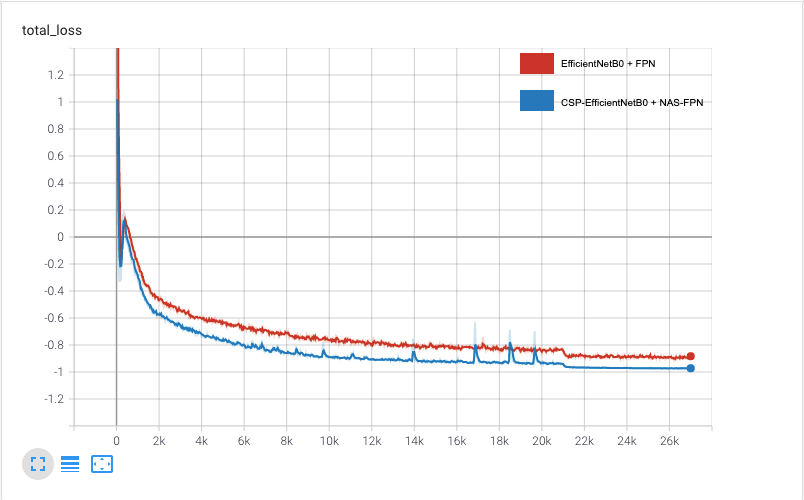}
    \caption{Comparing training loss plots of vanilla MaskRCNN\cite{he2018mask} with EfficientNetB0+FPN vs our CSP-EfficientNetB0+NAS-FPN.}
    \label{fig:lossplot}
\end{figure}

\begin{table}[!h]
\centering
\caption{Performance of Our model in comparison with comparable existing MaskRCNN\cite{he2018mask} model architectures on our primary dataset.}
\label{table:5}
\begin{tabular}{ |c|c|c|c|c| } 
 \hline
 \textbf{Work} & \textbf{Model} & \textbf{BoxAP} & \textbf{MaskAP} & \textbf{mIOU}  \\
 \hline
        & ResNet50-FPN & 62.4 & 63.3 & 0.82\\ 
 Prior  & SpineNet-49 & 63.3 & 64.1 & 0.84\\ 
        & EfficientNetB0-FPN & 63.4 & 64.3 & 0.84\\ 
 
\hline
 \textbf{Our Model} & \textbf{CSP-EfficientNetB0 + NAS-FPN + Dice Loss }& \textbf{64.7} & \textbf{65.6} & \textbf{0.86} \\ 
 \hline

\end{tabular}
\end{table}

\begin{figure}[!h]
\centering
     \begin{subfigure}[b]{0.45\textwidth}
         \centering
         \includegraphics[width=\textwidth]{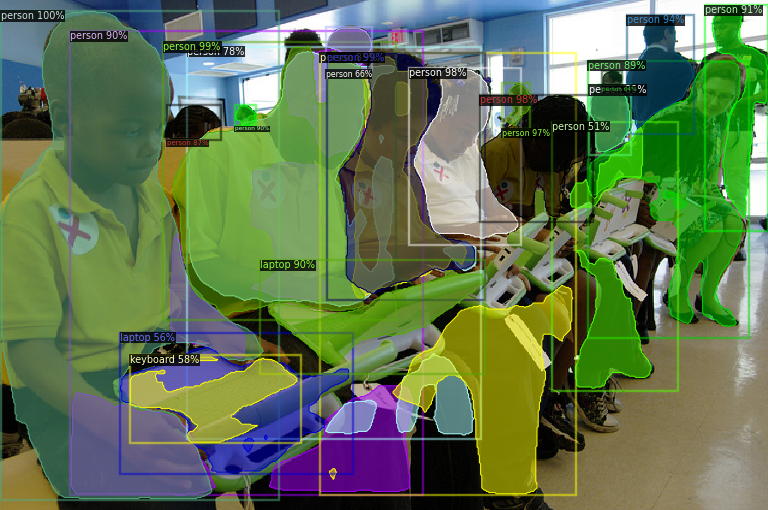}
         \caption{}
         \label{fig:s1}
     \end{subfigure}
     \hfill
     \begin{subfigure}[b]{0.45\textwidth}
         \centering
         \includegraphics[width=\textwidth]{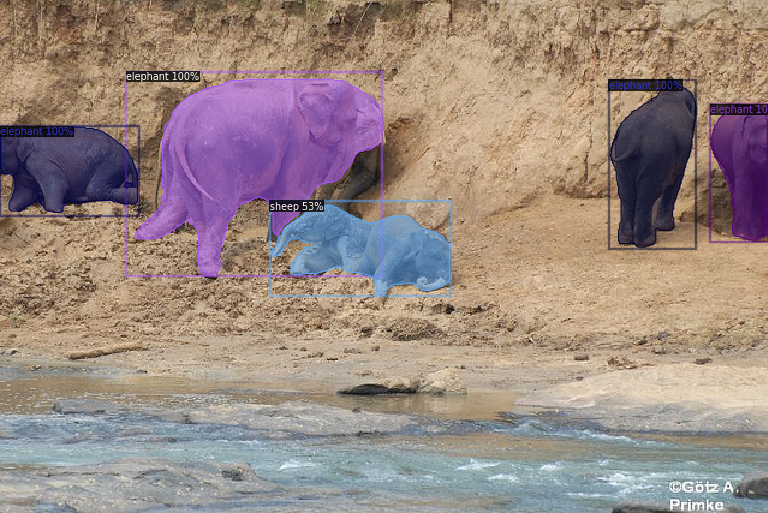}
         \caption{}
         \label{fig:s2}
     \end{subfigure}
     \hfill
     \begin{subfigure}[b]{0.45\textwidth}
         \centering
         \includegraphics[width=\textwidth]{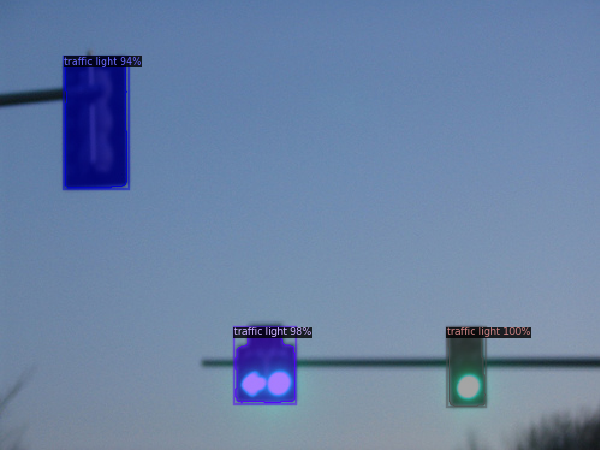}
         \caption{}
         \label{fig:s3}
     \end{subfigure}
     \hfill
     \begin{subfigure}[b]{0.45\textwidth}
         \centering
         \includegraphics[width=\textwidth]{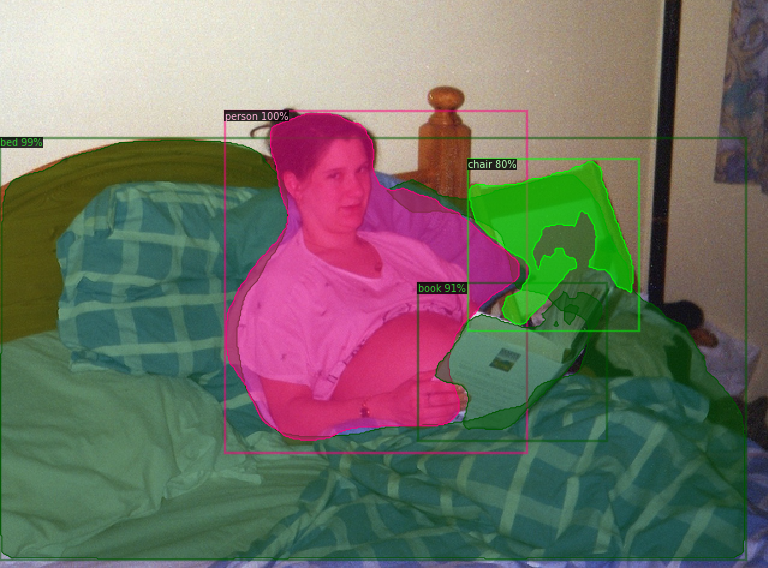}
         \caption{}
         \label{fig:s4}
     \end{subfigure}
     \hfill
     \begin{subfigure}[b]{0.45\textwidth}
         \centering
         \includegraphics[width=\textwidth]{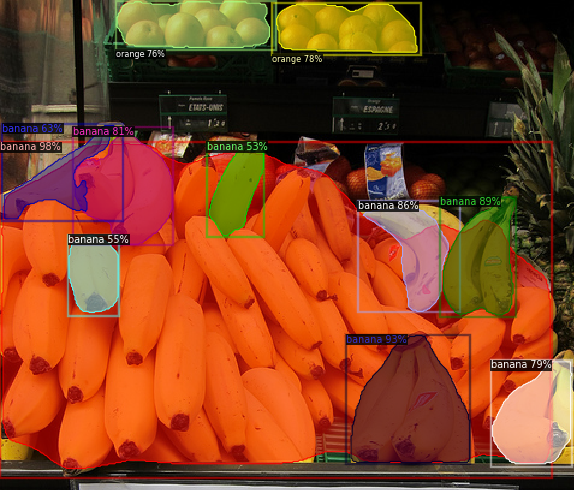}
         \caption{}
         \label{fig:s4}
     \end{subfigure}
     \hfill
     \begin{subfigure}[b]{0.45\textwidth}
         \centering
         \includegraphics[width=\textwidth]{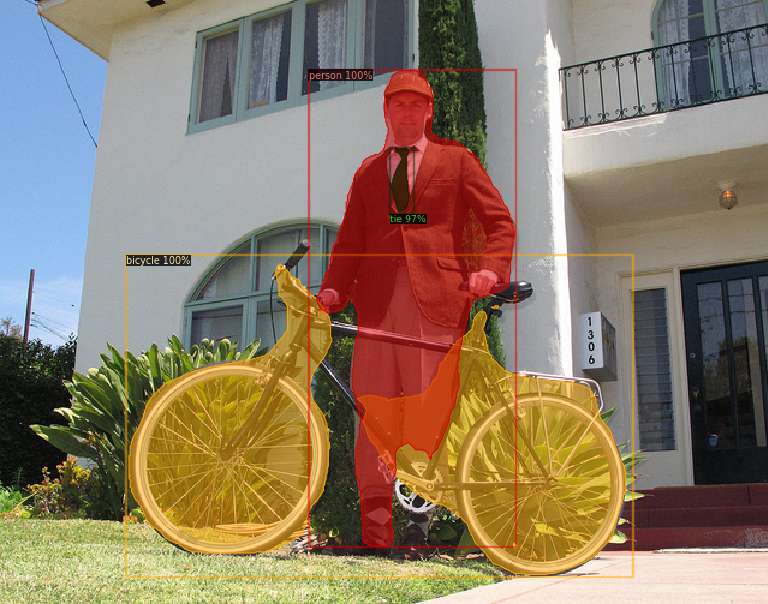}
         \caption{}
         \label{fig:s4}
     \end{subfigure}
\caption{Sample Detections on COCO minival \cite{lin2014microsoft} by our model.}
\label{fig:detections}
\end{figure}

\begin{figure}[!h]
\centering
     \begin{subfigure}[b]{0.45\textwidth}
         \centering
         \includegraphics[width=\textwidth]{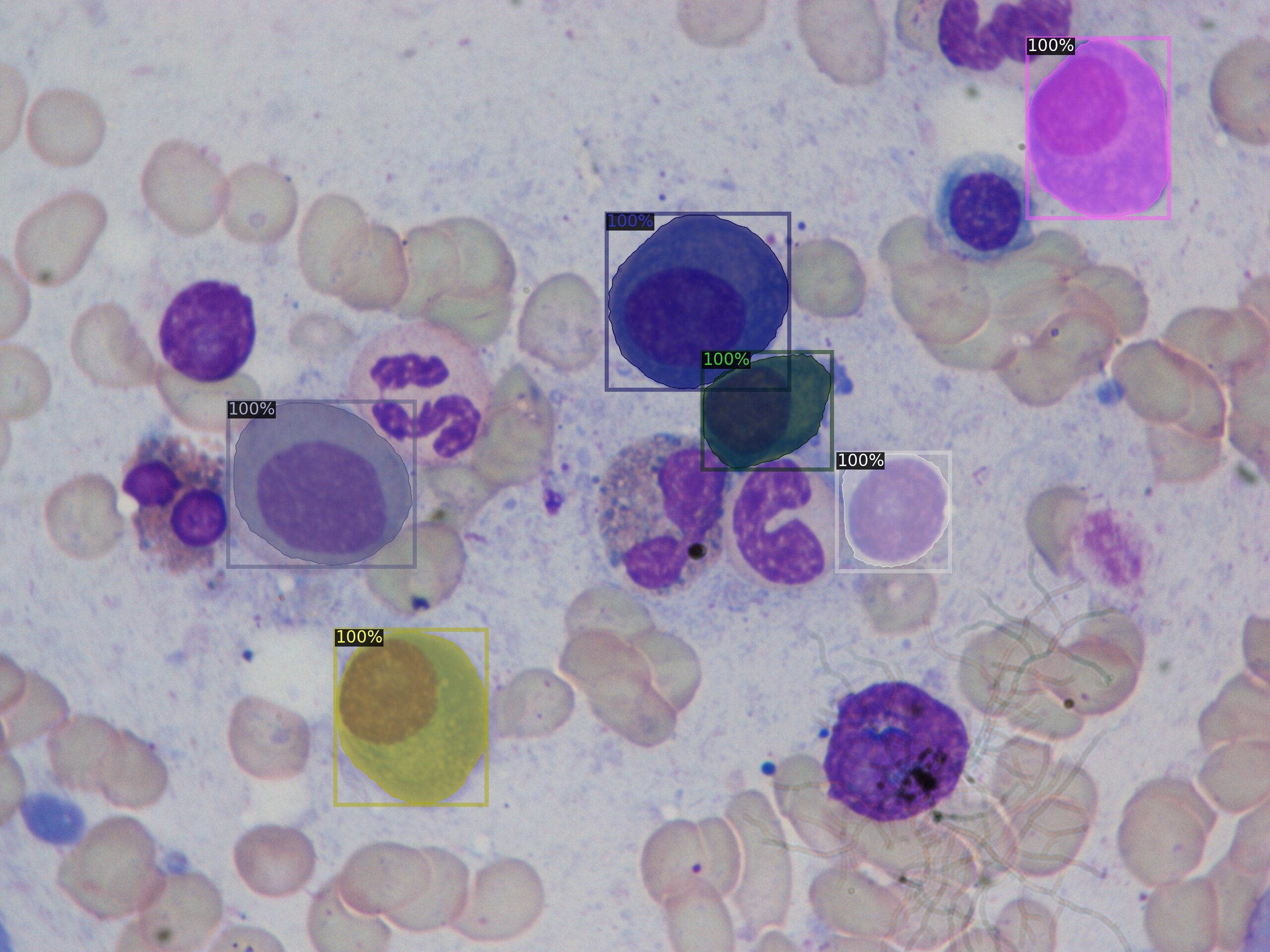}
         \caption{}
         \label{fig:s1}
     \end{subfigure}
     \hfill
     \begin{subfigure}[b]{0.45\textwidth}
         \centering
         \includegraphics[width=\textwidth]{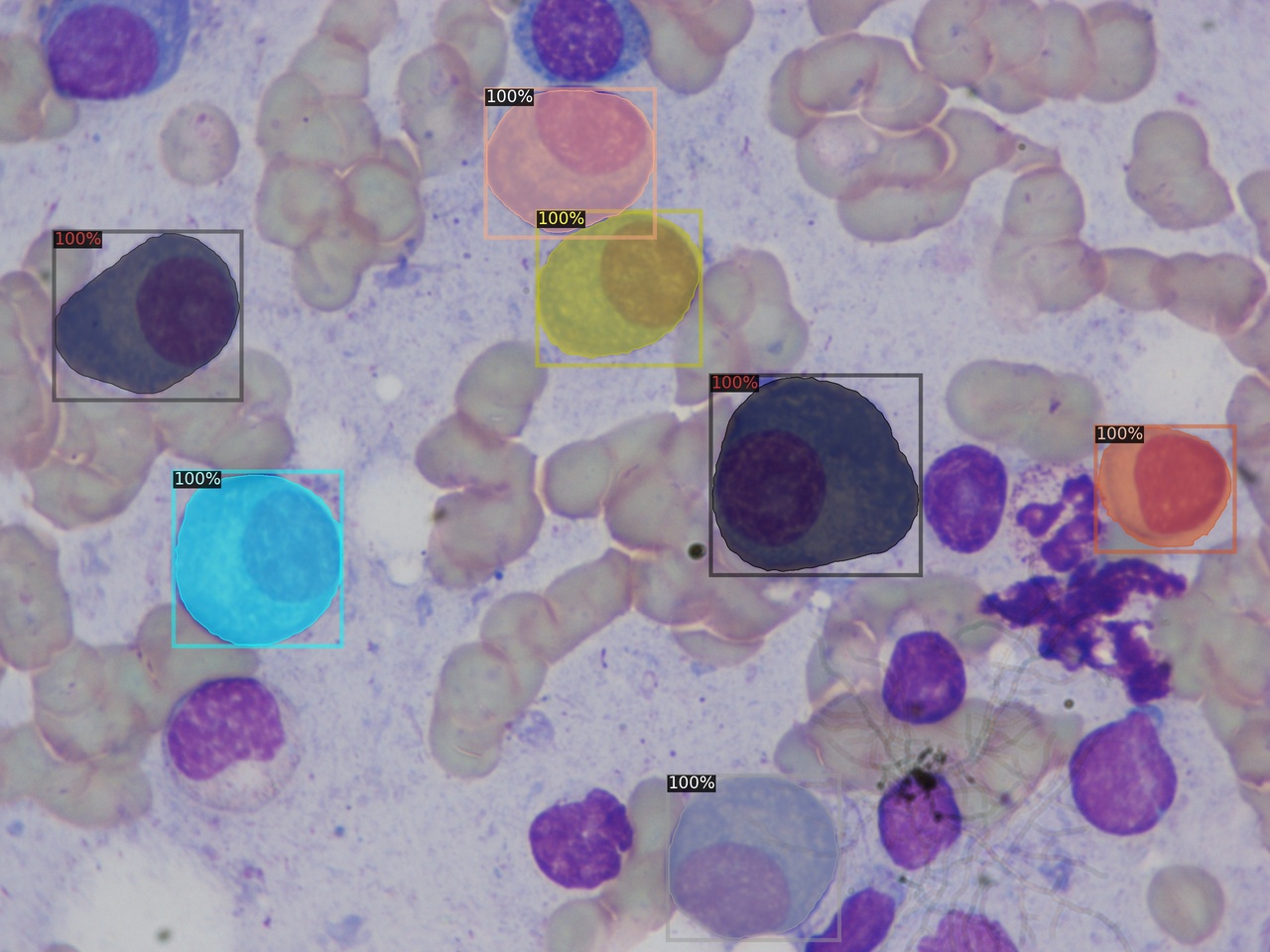}
         \caption{}
         \label{fig:s2}
     \end{subfigure}
     \hfill
     \begin{subfigure}[b]{0.45\textwidth}
         \centering
         \includegraphics[width=\textwidth]{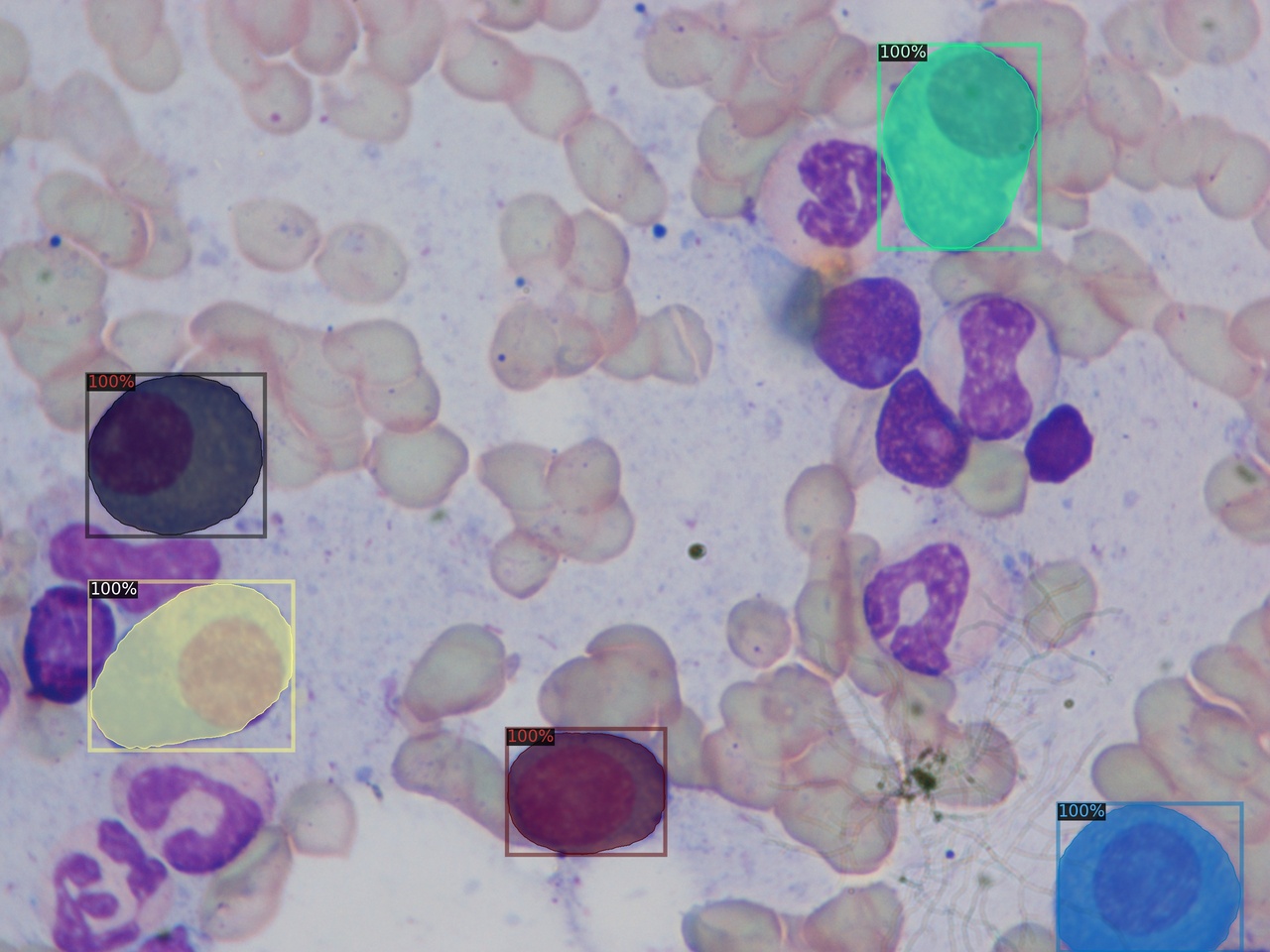}
         \caption{}
         \label{fig:s3}
     \end{subfigure}
     \hfill
     \begin{subfigure}[b]{0.45\textwidth}
         \centering
         \includegraphics[width=\textwidth]{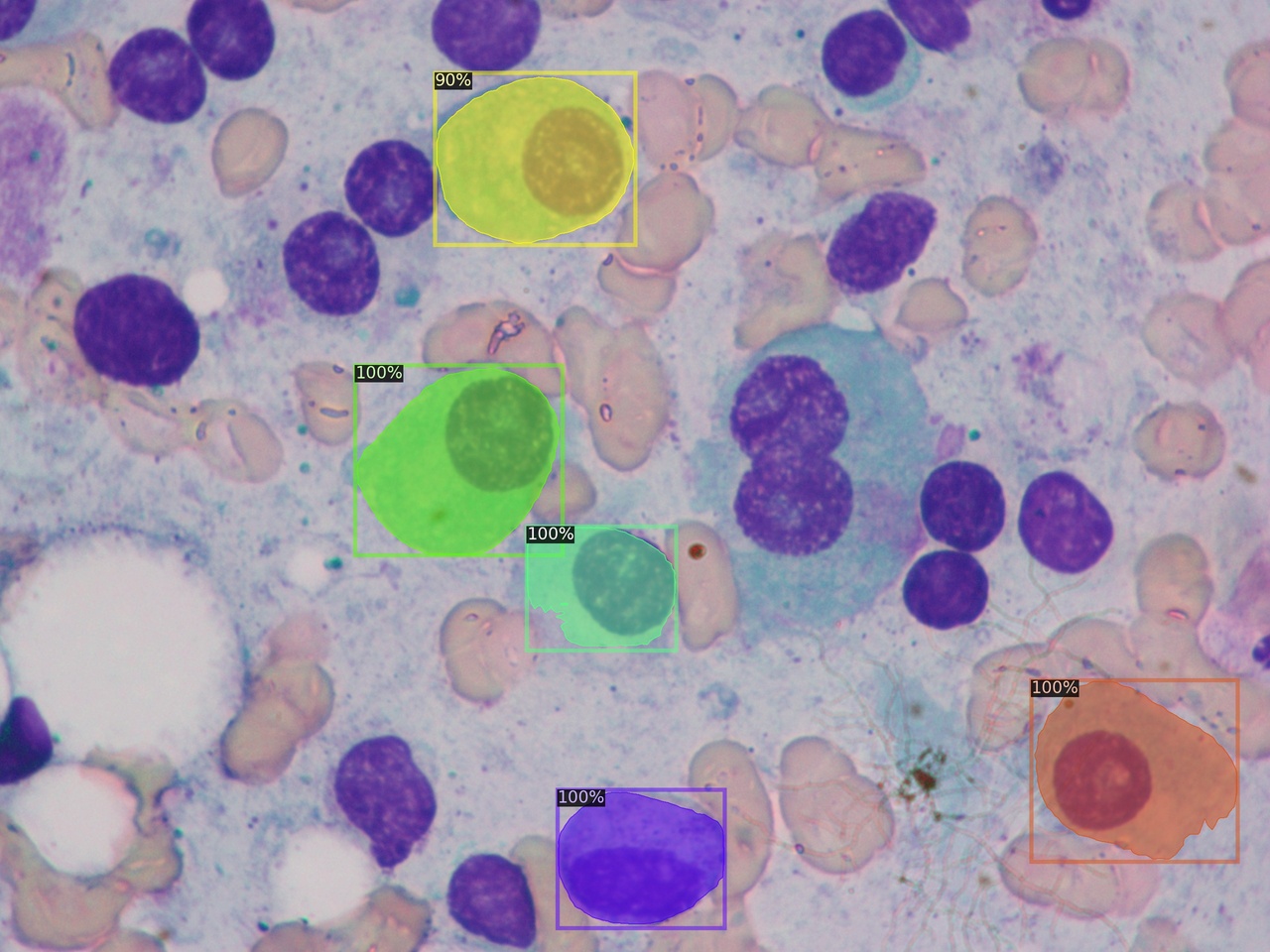}
         \caption{}
         \label{fig:s4}
     \end{subfigure}
     \hfill
     \begin{subfigure}[b]{0.45\textwidth}
         \centering
         \includegraphics[width=\textwidth]{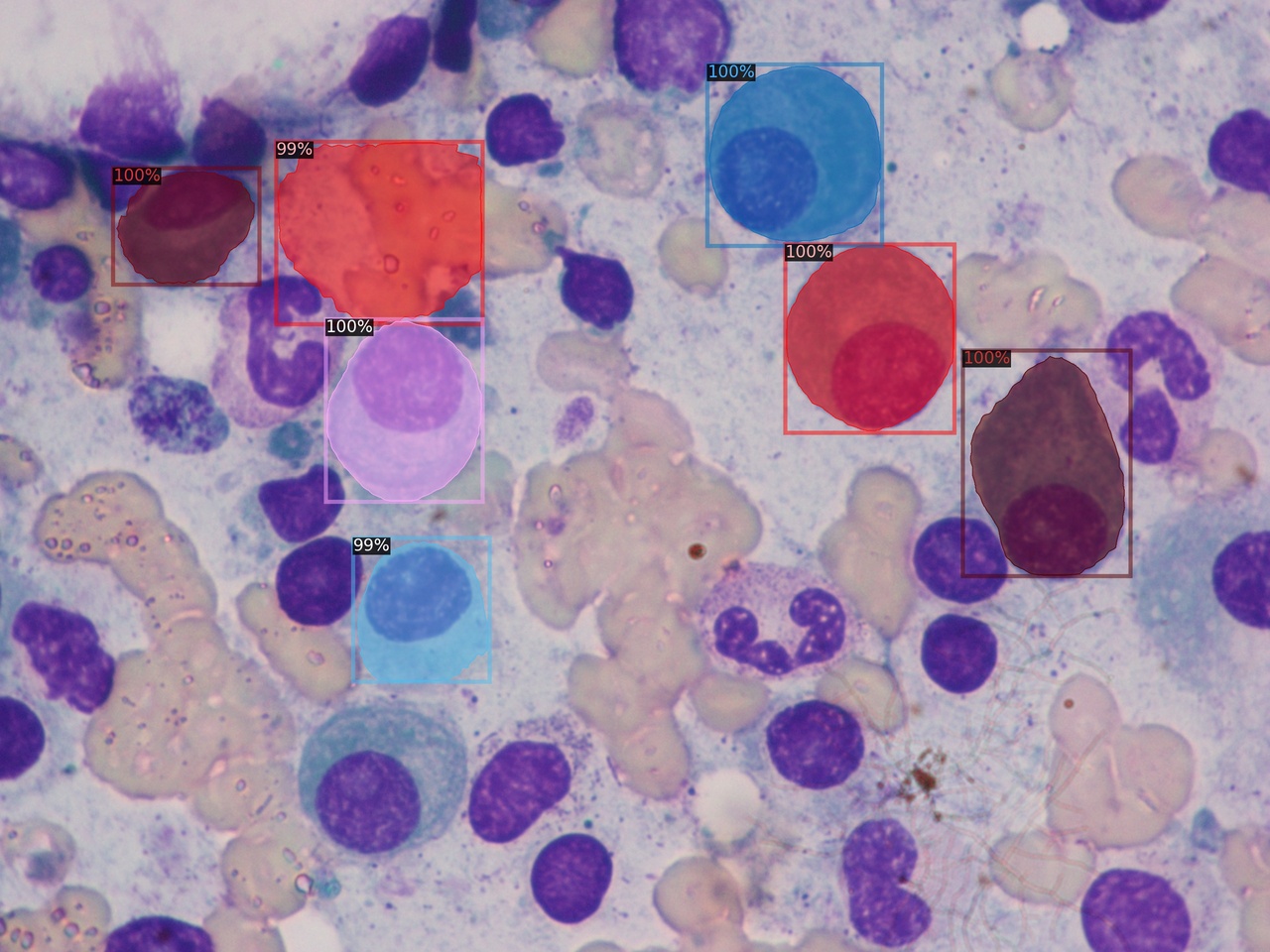}
         \caption{}
         \label{fig:s4}
     \end{subfigure}
     \hfill
     \begin{subfigure}[b]{0.45\textwidth}
         \centering
         \includegraphics[width=\textwidth]{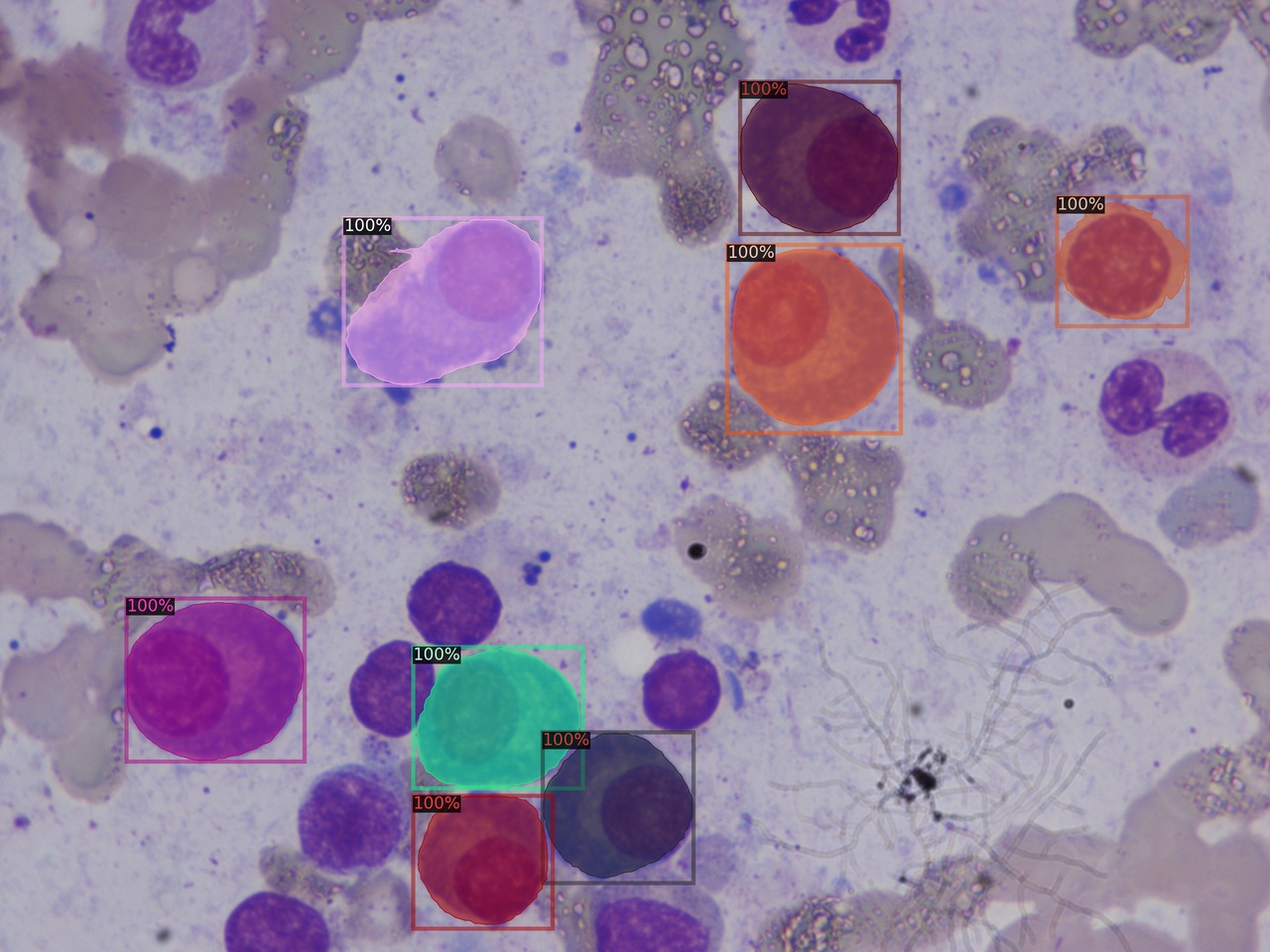}
         \caption{}
         \label{fig:s4}
     \end{subfigure}
\caption{Sample Detections on our primary dataset by our model.}
\label{fig:detections}
\end{figure}

\chapter{Conclusion and Future Work}\label{chapter:Conclusion and Future Work}
We find that our CSP-EfficientNet family of backbone architectures provide increased performance from the base vanilla EfficientNet\cite{tan2020efficientnet} Family of Models. We also find that NAS-FPN\cite{ghiasi2019nasfpn} also provides an increment in performance in comparision to other features networks. And our Modified loss objective for the mask head using dice coefficient\cite{Dice_1945} helps us achieve better IoU of predicted and ground truth masks and in result a better MaskAP.

Our model even on the base backbone architecture(CSP-EfficientNetB0) provides us with competitive performance on both the benchmark datasets for Object detection and instance segmentation. Also, it gives a competitive performance on our primary dataset.

We plan to further improve the performance of our model by finding more novel techniques and architectures to help us deploy a much accurate Multiple Myeloma cancer detection system at AIIMS, Delhi to help doctors promptly detect and diagnose Multiple Myeloma cancer patients.
\printbibliography

\end{document}

%% file: btech_thesis_cover_dikshant.tex
\def\addrone{Your address}
\def\addrtwo{Your city}

\def\degree{B.Tech. in Computer Science and Engineering}

\def\btptrack{Research}

\def\submissiondate{May 11, 2021}

\def\supervisorone{Dr.Anubha Gupta}

\def\supervisortwo{Shubham Goswami(PhD)}

\thispagestyle{empty}

\vspace{5.65in}

\begin{center}

\vspace{5.65in}

{\LARGE \bf { Multiple Myeloma Cancer Cell Instance Segmentation } }  
 \vspace{.3in}
 
 {\Large{Student Name: Dikshant Sagar}} \\  
 {\large{Roll Number: 2017338}} \\  
 \vspace{.1in} 


    \vspace{.65in}


\vspace{.65in}

 {BTP report submitted in partial fulfillment of the requirements 
\\for the Degree of B.Tech. in Computer Science \& Social Sciences}


on \submissiondate\\

\vspace{.65in}

\textbf{BTP Track}: \btptrack\\
\quad\\

  {\textbf{BTP Advisor} \\ 
\supervisorone              \\ 
\supervisortwo \\  }
\vspace{3.0in}

{Indraprastha Institute of Information Technology\\
New Delhi}

\end{center}


%% file: ref.bib
@misc{tan2020efficientnet,
      title={EfficientNet: Rethinking Model Scaling for Convolutional Neural Networks}, 
      author={Mingxing Tan and Quoc V. Le},
      year={2020},
      eprint={1905.11946},
      archivePrefix={arXiv},
      primaryClass={cs.LG}
}

@misc{tan2020efficientdet,
      title={EfficientDet: Scalable and Efficient Object Detection}, 
      author={Mingxing Tan and Ruoming Pang and Quoc V. Le},
      year={2020},
      eprint={1911.09070},
      archivePrefix={arXiv},
      primaryClass={cs.CV}
}

@misc{lin2018focal,
      title={Focal Loss for Dense Object Detection}, 
      author={Tsung-Yi Lin and Priya Goyal and Ross Girshick and Kaiming He and Piotr Dollár},
      year={2018},
      eprint={1708.02002},
      archivePrefix={arXiv},
      primaryClass={cs.CV}
}

@misc{wang2019cspnet,
      title={CSPNet: A New Backbone that can Enhance Learning Capability of CNN}, 
      author={Chien-Yao Wang and Hong-Yuan Mark Liao and I-Hau Yeh and Yueh-Hua Wu and Ping-Yang Chen and Jun-Wei Hsieh},
      year={2019},
      eprint={1911.11929},
      archivePrefix={arXiv},
      primaryClass={cs.CV}
}

@misc{du2020spinenet,
      title={SpineNet: Learning Scale-Permuted Backbone for Recognition and Localization}, 
      author={Xianzhi Du and Tsung-Yi Lin and Pengchong Jin and Golnaz Ghiasi and Mingxing Tan and Yin Cui and Quoc V. Le and Xiaodan Song},
      year={2020},
      eprint={1912.05027},
      archivePrefix={arXiv},
      primaryClass={cs.CV}
}

@misc{he2015deep,
      title={Deep Residual Learning for Image Recognition}, 
      author={Kaiming He and Xiangyu Zhang and Shaoqing Ren and Jian Sun},
      year={2015},
      eprint={1512.03385},
      archivePrefix={arXiv},
      primaryClass={cs.CV}
}

@misc{xie2017aggregated,
      title={Aggregated Residual Transformations for Deep Neural Networks}, 
      author={Saining Xie and Ross Girshick and Piotr Dollár and Zhuowen Tu and Kaiming He},
      year={2017},
      eprint={1611.05431},
      archivePrefix={arXiv},
      primaryClass={cs.CV}
}

@misc{huang2018densely,
      title={Densely Connected Convolutional Networks}, 
      author={Gao Huang and Zhuang Liu and Laurens van der Maaten and Kilian Q. Weinberger},
      year={2018},
      eprint={1608.06993},
      archivePrefix={arXiv},
      primaryClass={cs.CV}
}

@misc{russakovsky2015imagenet,
      title={ImageNet Large Scale Visual Recognition Challenge}, 
      author={Olga Russakovsky and Jia Deng and Hao Su and Jonathan Krause and Sanjeev Satheesh and Sean Ma and Zhiheng Huang and Andrej Karpathy and Aditya Khosla and Michael Bernstein and Alexander C. Berg and Li Fei-Fei},
      year={2015},
      eprint={1409.0575},
      archivePrefix={arXiv},
      primaryClass={cs.CV}
}

@misc{redmon2018yolov3,
      title={YOLOv3: An Incremental Improvement}, 
      author={Joseph Redmon and Ali Farhadi},
      year={2018},
      eprint={1804.02767},
      archivePrefix={arXiv},
      primaryClass={cs.CV}
}

@misc{lin2017feature,
      title={Feature Pyramid Networks for Object Detection}, 
      author={Tsung-Yi Lin and Piotr Dollár and Ross Girshick and Kaiming He and Bharath Hariharan and Serge Belongie},
      year={2017},
      eprint={1612.03144},
      archivePrefix={arXiv},
      primaryClass={cs.CV}
}

@misc{chen2018encoderdecoder,
      title={Encoder-Decoder with Atrous Separable Convolution for Semantic Image Segmentation}, 
      author={Liang-Chieh Chen and Yukun Zhu and George Papandreou and Florian Schroff and Hartwig Adam},
      year={2018},
      eprint={1802.02611},
      archivePrefix={arXiv},
      primaryClass={cs.CV}
}

@misc{zoph2017neural,
      title={Neural Architecture Search with Reinforcement Learning}, 
      author={Barret Zoph and Quoc V. Le},
      year={2017},
      eprint={1611.01578},
      archivePrefix={arXiv},
      primaryClass={cs.LG}
}

@article{Jiao_2019,
   title={A Survey of Deep Learning-Based Object Detection},
   volume={7},
   ISSN={2169-3536},
   url={http://dx.doi.org/10.1109/ACCESS.2019.2939201},
   DOI={10.1109/access.2019.2939201},
   journal={IEEE Access},
   publisher={Institute of Electrical and Electronics Engineers (IEEE)},
   author={Jiao, Licheng and Zhang, Fan and Liu, Fang and Yang, Shuyuan and Li, Lingling and Feng, Zhixi and Qu, Rong},
   year={2019},
   pages={128837–128868}
}

@article{palumbo2014international,
  title={International Myeloma Working Group consensus statement for the management, treatment, and supportive care of patients with myeloma not eligible for standard autologous stem-cell transplantation},
  author={Palumbo, Antonio and Rajkumar, S Vincent and San Miguel, Jesus F and Larocca, Alessandra and Niesvizky, Ruben and Morgan, Gareth and Landgren, Ola and Hajek, Roman and Einsele, Hermann and Anderson, Kenneth C and others},
  journal={Journal of clinical oncology},
  volume={32},
  number={6},
  pages={587},
  year={2014},
  publisher={American Society of Clinical Oncology}
}

@article{gupta2018pcseg,
  title={PCSeg: Color model driven probabilistic multiphase level set based tool for plasma cell segmentation in multiple myeloma},
  author={Gupta, Anubha and Mallick, Pramit and Sharma, Ojaswa and Gupta, Ritu and Duggal, Rahul},
  journal={PloS one},
  volume={13},
  number={12},
  pages={e0207908},
  year={2018},
  publisher={Public Library of Science San Francisco, CA USA}
}

@inproceedings{lin2014microsoft,
  title={Microsoft coco: Common objects in context},
  author={Lin, Tsung-Yi and Maire, Michael and Belongie, Serge and Hays, James and Perona, Pietro and Ramanan, Deva and Doll{\'a}r, Piotr and Zitnick, C Lawrence},
  booktitle={European conference on computer vision},
  pages={740--755},
  year={2014},
  organization={Springer}
}

@misc{matlab-simulink, title={Image Segmentation}, url={https://in.mathworks.com/discovery/image-segmentation.html}, journal={MATLAB & Simulink}}

@misc{sandler2019mobilenetv2,
      title={MobileNetV2: Inverted Residuals and Linear Bottlenecks}, 
      author={Mark Sandler and Andrew Howard and Menglong Zhu and Andrey Zhmoginov and Liang-Chieh Chen},
      year={2019},
      eprint={1801.04381},
      archivePrefix={arXiv},
      primaryClass={cs.CV}
}

@misc{woo2018cbam,
      title={CBAM: Convolutional Block Attention Module}, 
      author={Sanghyun Woo and Jongchan Park and Joon-Young Lee and In So Kweon},
      year={2018},
      eprint={1807.06521},
      archivePrefix={arXiv},
      primaryClass={cs.CV}
}

@inproceedings{Krizhevsky2009LearningML,
  title={Learning Multiple Layers of Features from Tiny Images},
  author={A. Krizhevsky},
  year={2009}
}

@misc{ghiasi2019nasfpn,
      title={NAS-FPN: Learning Scalable Feature Pyramid Architecture for Object Detection}, 
      author={Golnaz Ghiasi and Tsung-Yi Lin and Ruoming Pang and Quoc V. Le},
      year={2019},
      eprint={1904.07392},
      archivePrefix={arXiv},
      primaryClass={cs.CV}
}

@misc{he2018mask,
      title={Mask R-CNN}, 
      author={Kaiming He and Georgia Gkioxari and Piotr Dollár and Ross Girshick},
      year={2018},
      eprint={1703.06870},
      archivePrefix={arXiv},
      primaryClass={cs.CV}
}

@misc{wu2019detectron2,
  author =       {Yuxin Wu and Alexander Kirillov and Francisco Massa and
                  Wan-Yen Lo and Ross Girshick},
  title =        {Detectron2},
  howpublished = {\url{https://github.com/facebookresearch/detectron2}},
  year =         {2019}
}

@article{Dice_1945, title={Measures of the Amount of Ecologic Association Between Species}, volume={26}, ISSN={1939-9170}, DOI={https://doi.org/10.2307/1932409}, number={3}, journal={Ecology}, author={Dice, Lee R.}, year={1945}, pages={297–302} }
